\documentclass[]{aa}  
\usepackage{graphicx}
\usepackage{txfonts}
\usepackage{amssymb,natbib}

\def\vqsqusgr{\mbox{V4641 \mbox{Sgr}}}

\def\xtejodh{\mbox{XTE J1118+480}}

\def\xtejdsv{\mbox{XTE J1720-318}}

\def\cmmoinsdeux{\mbox{ cm}^{-2}}

\def\microns{\mbox{ } \mu \mbox{m}}

\def\mags{\mbox{ magnitudes}}

\def\hour{^{h}}

\def\min{^{m}}

\def\secp{{\rlap.}^{s}}

\def\deg{^{\circ}}

\def\amin{^\prime}

\def\asecp{{\rlap.}^{\prime \prime}}

\def\Av{A_{\rm v}}

\def\nh{N_{\rm H}}

\def\ltsima{\; \buildrel < \over \sim \;}
\def\simlt{\lower.5ex\hbox{\ltsima}}            
\def\gtsima{\; \buildrel > \over \sim \;}
\def\simgt{\lower.5ex\hbox{\gtsima}}            

\begin{document}
   \title{Optical/near-infrared observations of the black hole candidate
 $\xtejdsv$: from high-soft to low-hard state\thanks{Based on ESO observations through programme \# 070.D-0340}}


\author{S. Chaty\inst{1} \and N. Bessolaz\inst{2}}

\offprints{S. Chaty}

\institute{AIM - Astrophysique Interactions Multi-\'echelles 
(UMR 7158 CEA/CNRS/Universit\'e Paris 7 Denis Diderot), CEA Saclay,
DSM/DAPNIA/Service d'Astrophysique, B\^at. 709, L'Orme des Merisiers,
FR-91191 Gif-sur-Yvette Cedex, France.
\email{chaty@cea.fr}
\and 
Laboratoire d'Astrophysique, Observatoire de Grenoble, 
BP 53, FR-38041 Grenoble Cedex 9, France}

   \date{Received; accepted }

 
  \abstract
   {}
   {To gain a better understanding of high-energy Galactic sources,
we observed the Galactic X-ray  binary and black hole candidate $\xtejdsv$
in the optical  and  near-infrared, 
just  after the onset  of its X-ray outburst in January, 2003.
These observations were obtained with the ESO/NTT as the Target of Opportunity,
in February and April 2003.}
   {We performed an accurate astrometry and analysed photometrical  and
spectroscopic observations. We then produced a colour-magnitude
diagram, looked at the overall evolution of the multi-wavelength light curve,
 and analysed the  spectral energy distribution.}
   {We  discovered the optical counterpart in the R-band (R $\sim 21.5$) and
confirmed the near-infrared counterpart. 
We show that, for an absorption between 6 and 8 magnitudes, 
$\xtejdsv$ is  likely to be an intermediate  mass X-ray binary 
located  at a distance  between 3  and  10 kpc,  hosting  a  main
sequence star  of spectral type between  late B and early  G.
Our second set of observations took place simultaneously with the third
 secondary outburst present in X-ray and near-infrared light curves.
 The evolution  of its spectral  energy distribution 
shows that $\xtejdsv$ entered a transition from a high-soft to a low-hard
state in-between the two observations.}
   {}


\authorrunning{S. Chaty \& N. Bessolaz}
\titlerunning{Optical/NIR observations of the black hole candidate $\xtejdsv$}

   \maketitle
%

\section{Introduction}

X-ray binaries are constituted of a compact object and a companion
star, the former attracting matter from the later, either through
an accretion disc or the wind. They are usually divided in 2 sub-classes:
high mass X-ray binaries and low mass X-ray binaries, hosting early-type
and late-type stars, respectively. Since accretion and ejection phenomena
usually occur in these objects, they are ideal laboratories
for studying relativistic phenomena and the formation
and evolution of compact objects in binaries.
However, to study them, we first have to derive 
the important parameters related to
the nature of these systems, i.e. the distance, nature of the compact object, 
spectral type of the companion star, type of accretion, orbital parameters, 
etc.  Because of the way
they are formed, most of the observed Galactic X-ray binaries are located
in the Galactic plane or even towards the Galactic centre, 
and therefore associated with very high
absorption (up to $\Av \sim 50 \mags$) 
because of the presence of gas and dust
in this region. In this case, near-infrared (NIR) observations
prove to be particularly useful, since the radiation is less absorbed
at NIR wavelengths than at optical ones (see, e.g., \citeauthor{chaty:2002b} 
\citeyear{chaty:2002b}).
Furthermore,  X-ray binaries have to be studied in a multi-wavelength context
to disentangle all parts of the system emitting at various 
wavelengths: the accretion disc from high-energies to NIR, 
the companion star from
ultra-violet to NIR, the jets from radio to X-rays, etc. 
(see, e.g., \citeauthor{chaty:2005} \citeyear{chaty:2005}).

On January 9, 2003, the All Sky Monitor ({\it ASM}) of the {\it Rossi-XTE} 
satellite discovered a new source in the X-ray sky: $\xtejdsv$, 
in the direction of the Galactic bulge 
(\citeauthor{remillard:2003} \citeyear{remillard:2003}).
The 2-12 keV flux was initially $\sim$ 130
mCrab and reached $\sim$ 430 mCrab on January 10, 2003. Spectroscopic
observations with {\it XMM-Newton} were carried out on  February 20, 2003,
allowing to estimate the column density of hydrogen on the line of sight:
$\nh=1.33 \times 10^{22} \cmmoinsdeux$ (\citeauthor{markwardt:2003a} \citeyear{markwardt:2003a}). An iron line was detected
at 6.2 keV with 95 eV equivalent width, and no low or high frequency
oscillation was detected (\citeauthor{markwardt:2003a} \citeyear{markwardt:2003a}). 
The 2-10 keV flux was estimated to be
$1.6 \times 10^{-9} \mbox{ erg cm}^{-2} \mbox{s}^{-1}$ 
(\citeauthor{gonzalez-riestra:2003} \citeyear{gonzalez-riestra:2003}).
The source was not detected during {\it INTEGRAL}/{\it IBIS} 
observations on February 28, 2003, but became visible at the end of the burst 
during {\it IBIS} surveys of the Galactic centre from
the end of March 2003 (\citeauthor{goldoni:2003} \citeyear{goldoni:2003}). The source 
reached $\sim$ 25 mCrab in the 15-40 keV band and 
became detected in the 40-100 keV energy band at $\sim$ 30 mCrab on 
 April 6 and 7, 2003, indicating that the source had undergone a change of state,
as suggested by \citeauthor{goldoni:2003} (\citeyear{goldoni:2003}).
The high-energy observations suggest that the compact object
is a black hole, first because of its early
spectral evolution that was very similar to black hole X-ray transients
(\citeauthor{remillard:2003} \citeyear{remillard:2003}), and also because of X-ray spectral parameters 
(\citeauthor{cadolle-bel:2004} \citeyear{cadolle-bel:2004}) and the presence of an iron line (\citeauthor{markwardt:2003a} \citeyear{markwardt:2003a}).

A radio counterpart was discovered with the VLA on January 15, 2003, and
confirmed with ATCA on January 16, 2003: only one
radio source included in the {\it Rossi-XTE} error box significantly varied 
from 0.32 $\pm$ 0.04 to 4.9 $\pm$ 0.1 mJy at 4.9 GHz (\citeauthor{rupen:2003} \citeyear{rupen:2003}).  
Radio observations took place from January to August 2003, allowing us to study
the correlation between radio and X-ray fluxes (\citeauthor{brocksopp:2005} \citeyear{brocksopp:2005}).  

A NIR counterpart was discovered then by \citeauthor{nagata:2003} (\citeyear{nagata:2003}) on
 January 18, 2003. Thirteen observations in $J$, $H$, and $K_{s}$,
until  May 21, 2003, allowed to measure the exponential decay 
following the burst, which was equal to 60 days (\citeauthor{nagata:2003} \citeyear{nagata:2003}).

In this paper, we will first describe our optical and NIR observations
and data reduction in Sect. \ref{obs}, then report on our astrometry, 
photometry, and spectroscopy results in Sect. \ref{results}.
We will then focus on constraining the companion star spectral
type in Sect. \ref{companion}, and finally we will analyse the evolution
of the $\xtejdsv$ light curve and 
spectral energy distribution (SED) in Sect. \ref{SED}.


\section{Observations and data reduction} \label{obs}

Our observations were carried out as part of the Target of Opportunity
(ToO) programme 070.D-0340 (PI: S. Chaty) dedicated to the study of new
Galactic high-energy sources and jet sources. They were triggered so as 
to be conducted at the same time as {\it INTEGRAL} ToO
observations, and we asked for two periods of observations.  
The first set of observations took place on  February 28, 2003, 
and the second one on  April 24, 2003. 
On  February 28, 2003, we obtained NIR photometry in $J$-, $H$-,
and $K_{s}$-bands with the spectra-imager SofI, and optical photometry
in $B$-, $V$-, $R$-, and $I$-bands with EMMI, both installed on the NTT. On 
April 24, 2003, in addition to optical and NIR photometry, we also carried
out NIR spectroscopy with SofI between 0.9 and $1.6 \microns$.
We used the large field imaging of SofI's detector, giving an image scale of
0.288" pixel$^{-1}$ and a field of view of $4.94\amin \times 4.94\amin$, 
and the EMMI detector with an image scale of 0.32" pixel$^{-1}$ 
and a binning $2 \times 2$, giving a field of view of 
$9.9\amin \times 9.1\amin$.

Concerning the NIR observations, we repeated one set of observations for each
filter with 9 different 30" offset positions, including $\xtejdsv$,
with an integration time of 90 seconds for each exposure, 
following the standard jitter procedure that allows us to cleanly subtract 
the blank NIR sky emission. We
observed two photometrical standard stars of the faint NIR standard star
catalogue of \citeauthor{persson:1998} (\citeyear{persson:1998}): sj9157 on  February 28, 2003, and sj9172
on  April 24, 2003. We also performed rapid photometry in the $K_{s}$-band 
for half an hour to detect rapid variations of
magnitude by taking a set of 90 images with 2 s integration time
each. We binned the images by three using a median filter, and we
carried out aperture photometry. 

Concerning the optical observations, we acquired 300 s exposures in each
filter, except for the B-band (200 s), using a 2x2 binning to
increase the sensitivity. We observed the standard star RU152 in R- and I-bands.
Since we did not have any standard star observations in the B
and V filters, we used mean zero-points taken from the EMMI 
website\footnotemark[1]. 

\footnotetext[1]{www.ls.eso.org/lasilla/sciops/ntt/emmi/} 

We used the Image Reduction and Analysis Facility \rm{(IRAF)} suite to
perform data reduction, carrying out standard procedures of optical
and NIR image reduction, including flat-fielding and NIR sky subtraction.
The zero-points we obtained are reported in Table \ref{zeropoints}.
As we had only one standard star observation
available for each
night, we used characteristic extinction coefficients at la Silla:
$ext_B=0.214$, $ext_V = 0.125$, $ext_R=0.091$, $ext_I= 0.051$, $ext_J
= 0.08$, $ext_H = 0.03$, and $ext_{Ks} = 0.05$,
to transform instrumental magnitudes into apparent magnitudes.
The observations were
performed through an airmass between 1 and 1.4.

Concerning the NIR spectroscopy, we took 12 spectra, half of them with
an offset of 30" from the other half, 
to subtract the blank NIR sky, giving a total integration
time of 180 s. To extract spectra and perform wavelength and flux
calibrations, we used the {\it IRAF noao.twodspec} package.  We used
the standard star sj9157 already mentioned above
to perform flux calibration. Since the
spectral type of this star is unknown, we used the calibrated $J$, $H$,
and $Ks$ magnitudes of \citeauthor{persson:1998} (\citeyear{persson:1998}) to deduce its spectral 
type by using a colour-colour diagram taken from \citeauthor{cox:2000} (\citeyear{cox:2000})
and assuming a main sequence star with negligible interstellar
absorption on the line of sight. We then synthesised a blackbody
spectrum with effective temperature, corresponding to the determined
spectral type. This flux calibration is quite crude, since NIR stellar spectra
 often present broad absorption features;
however, it gives a good indication of the flux. 

\begin{table}
\centering
\begin{tabular}{ccc}
\hline \hline
  & SofI observations & Emmi observations \\
\hline 
 February 28, 2003 & $Z_j=2.229\pm0.009$ &         \\
MJD 52698        & $Z_h=2.259\pm0.007$ &         \\
                 & $Z_{k_{s}}=2.862\pm0.006$ &         \\
\hline
 April 24, 2003 & $Z_j=2.268\pm0.007$ & $Z_b=25.27\pm0.03$   \\
MJD 52753     & $Z_h=2.451\pm0.009$ & $Z_v=25.98\pm0.01$  \\
              & $Z_{k_{s}}=2.999\pm0.011$ & $Z_r=26.21\pm0.02$    \\ 
	      &                           & $Z_i=25.57\pm0.03$    \\
\hline
\end{tabular}
\caption{Zero-points derived from SofI and EMMI observations. 
MJD = JD - 2400000.5}
\label{zeropoints}
\end{table} 

\section{Astrometry, photometry, and spectroscopy results} \label{results}

We used the $K_{s}$ image of the $\xtejdsv$ field taken on  January 21, 2003
(\citeauthor{obrien:2003} \citeyear{obrien:2003})
to identify $\xtejdsv$ in our NTT images. We then determined 
the position of the $\xtejdsv$ NIR counterpart by deriving the astrometric
solution, using $\sim 12$ stars taken from the GSC2 catalogue: the position we
measured was: $\alpha = 17\hour 19\min 58\secp 988 \pm0\secp 008$;
$\delta = -31\deg 45\amin 01\asecp 21 \pm 0\asecp15$ (equinox
J2000). This position is consistent with other determinations
(Table \ref{astrometry}).

We discovered the optical counterpart in the R- and I-bands at $\alpha
= 17\hour 19\min 58\secp994 \pm 0\secp007$, $\delta = -31\deg 45\amin
01\asecp46 \pm 0\asecp15$ (equinox J2000), a position that is consistent 
with the NIR counterpart. 
We present $BVRI$ magnitudes in Table \ref{mag_optir}. The R and I 
magnitudes are consistent with detection
limits given by \citeauthor{nagata:2003} (\citeyear{nagata:2003}): $R>18$ and $I>16.5$. 
We give a lower limit for the B- and V-bands, as we did not detect
any counterpart in these bands.

Since $\xtejdsv$ is located close to the Galactic centre, we had to
perform crowded field photometry to obtain precise NIR
magnitudes, using the {\it noao.daophot} package. This procedure,
described in \citeauthor{massey:1992} (\citeyear{massey:1992}), consists of creating an empirical
point-spread function with isolated bright stars, applying this model to the
whole field, cancelling the contributions of neighbour stars, measuring
the flux of the object itself, and then applying aperture correction
(due to the use of a smaller aperture for measuring $\xtejdsv$ magnitude
than for standard stars). 
This procedure allows us to get photometry with better than 1\% accuracy.  
We present
the apparent $J$, $H$, and $K_{s}$ magnitudes measured in  February and
April 2003 in Table \ref{mag_optir}. Uncertainties were determined from CCD
readout and signal noise. We note that we do not include  
the $H$-band observation of  February 28, 2003, 
here because of bad sky subtraction.
 
\begin{figure}
\resizebox{\hsize}{!}{\includegraphics[width=6cm, angle=90]{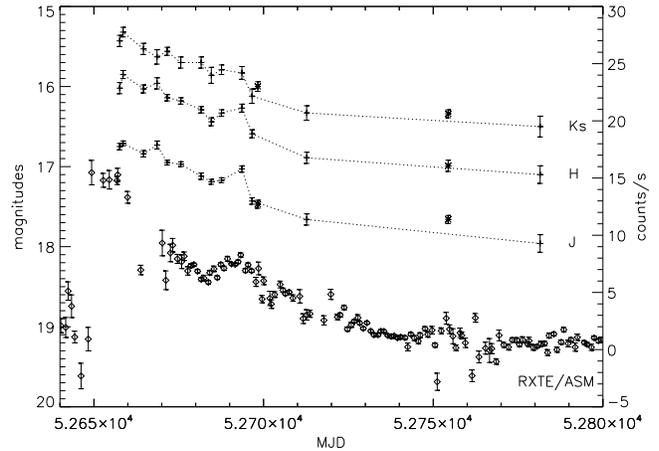}}
\caption{Multi-wavelength $\xtejdsv$ light curve.
From bottom to top: {\it Rossi-XTE} X-ray light curve (indicated
by diamonds); NIR J, H, and $K_{s}$ light curves, respectively.
NIR data taken from \citeauthor{nagata:2003} (\citeyear{nagata:2003}) are reported with '+', 
and data from this paper with '$\ast$'. MJD = JD - 2400000.5}
\label{IRlc}
\end{figure}

Our NIR photometrical observations are reported in Fig. \ref{IRlc},
where we also included results from \citeauthor{nagata:2003} (\citeyear{nagata:2003}). Our magnitudes,
indicated by '$\ast$', are consistent with measures from 
\citeauthor{nagata:2003} (\citeyear{nagata:2003}). 
We notice the similar behaviour of X-ray and NIR light curves,
particularly during the first maximum when the coverage is more complete, 
and we point out that we observed an increase in NIR on  April 24, 2003 
(MJD 52753),
observed by chance  at exactly the same time as an X-ray increase
seen on the light curve.

\begin{figure*}
\setlength{\unitlength}{1.0cm}
\begin{picture}(8.5,8.5)(1.,-7.)
\includegraphics{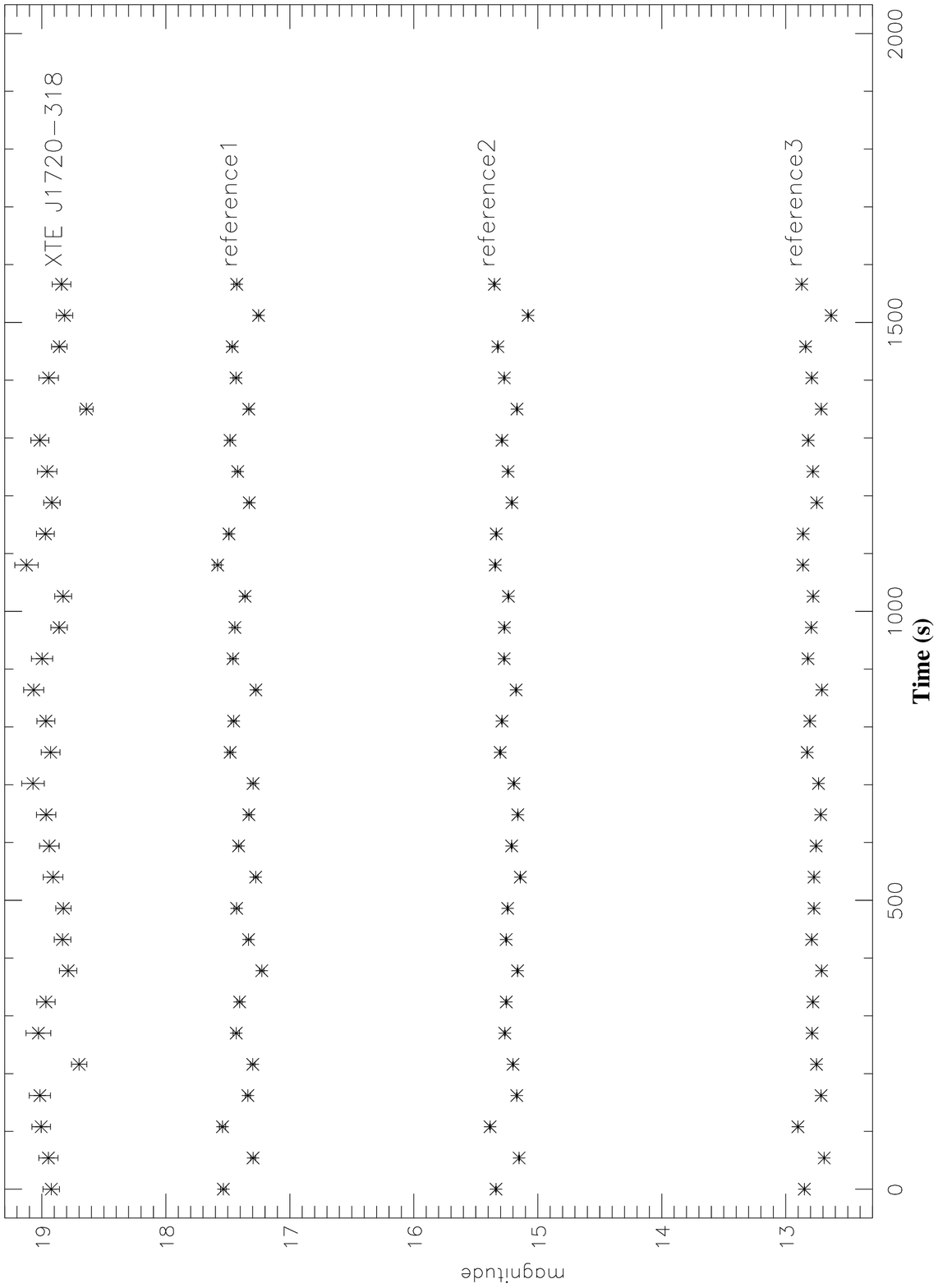}
\end{picture}
\begin{picture}(8.5,8.5)(0.5,-7.)
\includegraphics{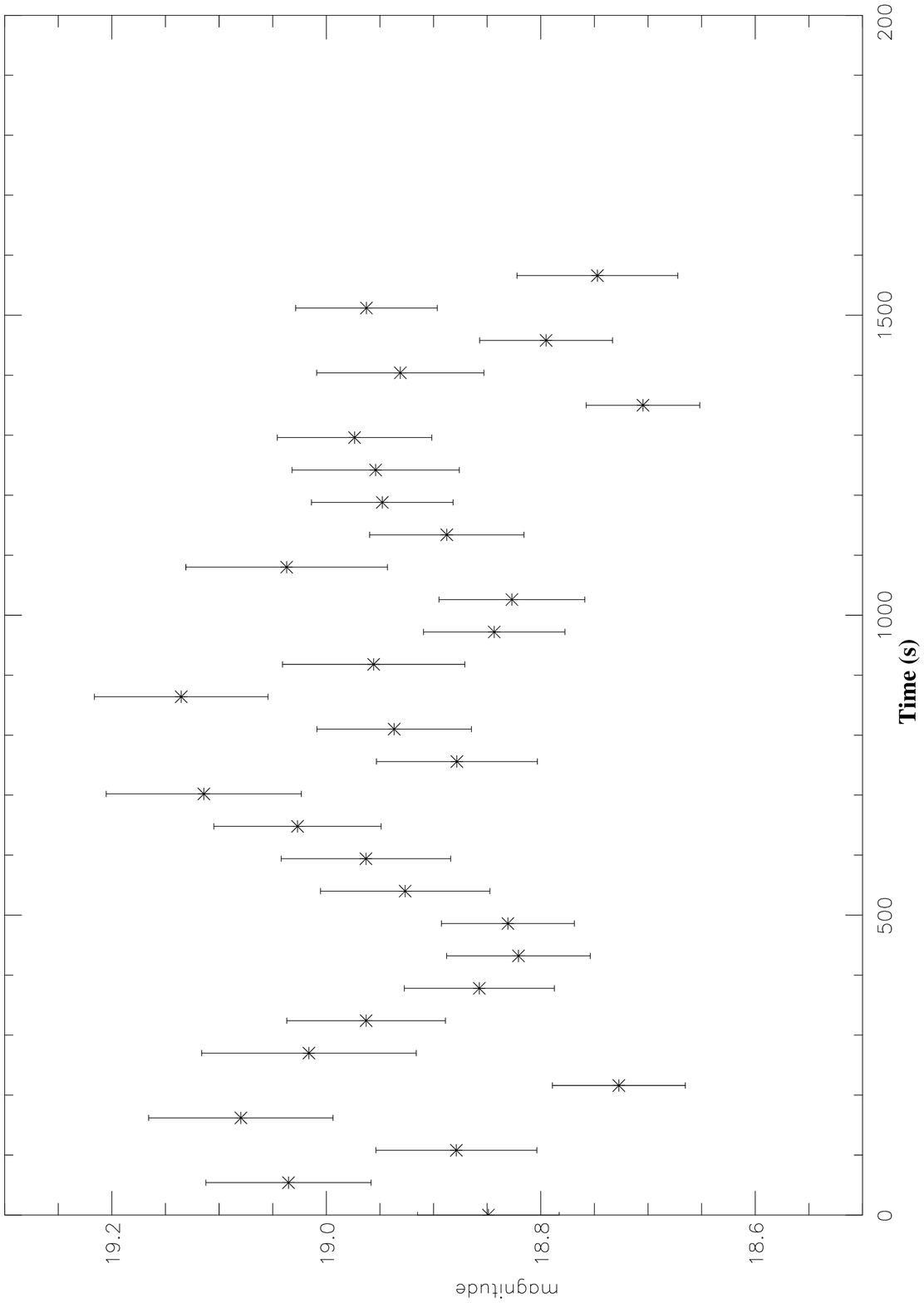}
\end{picture}
\caption[]{\label{rapidKs}    
Left panel: $K_{s}$ light curve of $\xtejdsv$ and three comparison stars
of the field.
Right panel: $K_{s}$ light curve of $\xtejdsv$ corrected from sky variations
(the y-axis is in instrumental magnitude).}
\end{figure*}

To analyse the rapid $K_{s}$ photometry, we performed aperture
photometry of $\xtejdsv$ and three other stars of
different magnitudes in the field. 
We show the results in the left panel of Fig.
\ref{rapidKs}: we see the same type of variability for $\xtejdsv$ as for
comparison stars, showing that this variability is more likely due 
to sky variations than to the X-ray binary itself.  
Dividing $\xtejdsv$ flux by the means of comparison star fluxes, 
we corrected the NIR light curve of $\xtejdsv$ from this sky variability, 
to detect $\xtejdsv$ intrinsic variations, if any.  The corrected 
light curve is presented in the right panel of Fig. \ref{rapidKs}. 
The mean instrumental $K_{s}$ magnitude, obtained by median-filtering all 
images, is $18.93\pm0.04 \mags$, and the standard deviation of all measures 
is $\sim 0.11 \mags$. 
We can see a periodicity of $\sim 600$ s in Fig. \ref{rapidKs};
however, since the interval of magnitude variation is around 0.3 ($3 \sigma$), 
it is not significant enough to interpret it as $\xtejdsv$ intrinsic
variations. Furthermore, a discrete Fourier analysis of 
these data does not give significant evidence of periodicity.
Absence of significant variability at the level of +/- 0.1 mag
is consistent with the similar analysis done earlier in the outburst
by \citeauthor{obrien:2003} (\citeyear{obrien:2003}).

\begin{figure}
\resizebox{\hsize}{!}{\includegraphics[width=6cm, angle=0]{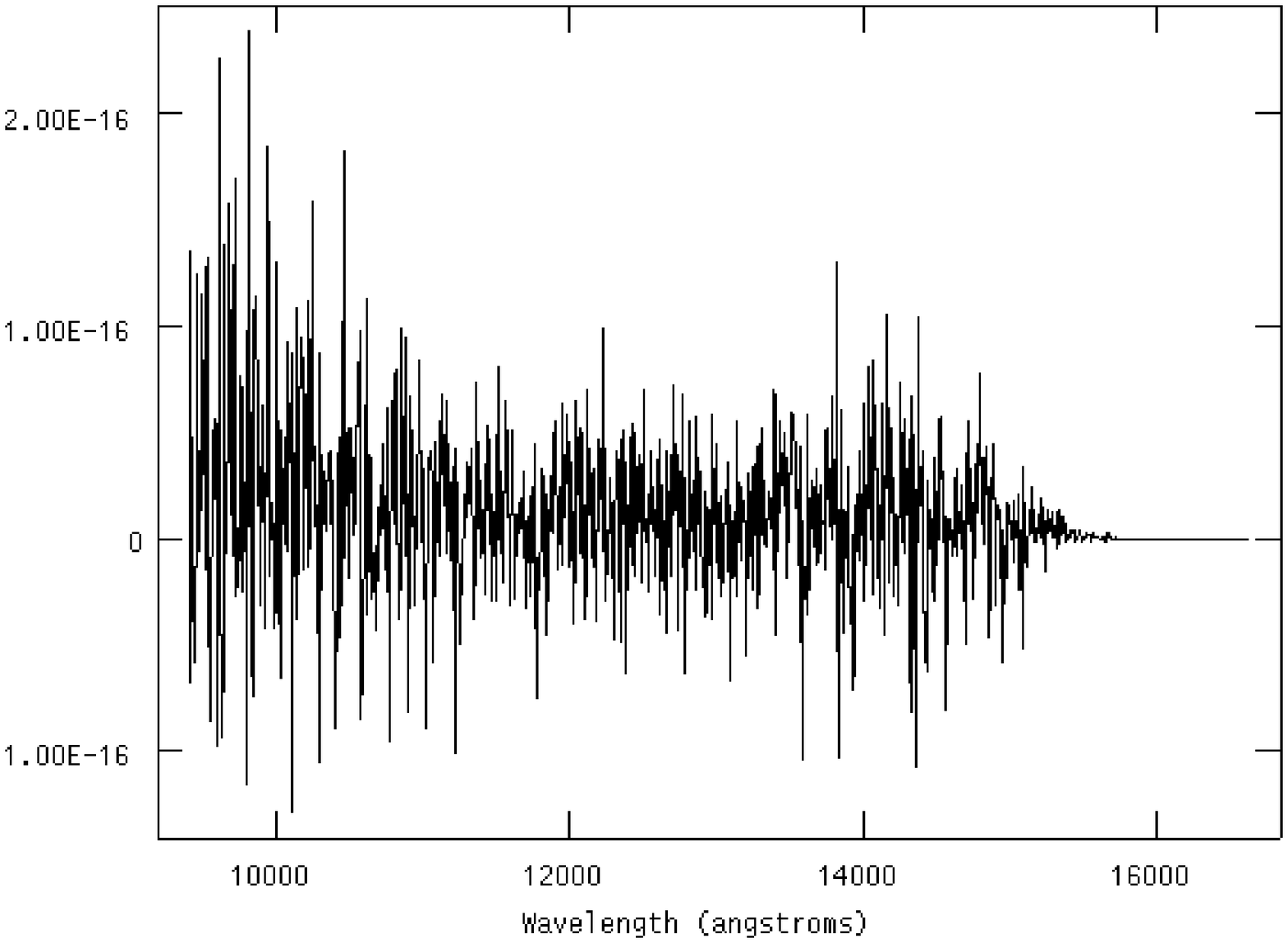}}  
\caption{NIR $K_{s}$-band spectrum of $\xtejdsv$. The flux is given in 
$\mbox{ erg cm}^{-2} \mbox{ s}^{-1} \AA^{-1}$.
\label{specXTEfluxcalib}}
\end{figure}

Finally, concerning the spectroscopy, we show 
the flux and wavelength calibrated spectrum we obtained 
in Fig. \ref{specXTEfluxcalib}.
Although the signal-to-noise ratio was too small to detect any line,
the spectrum allowed us to estimate an upper limit to $\xtejdsv$ flux: 
$\sim 2 \times 10^{-16} \mbox{ erg cm}^{-2} \mbox{ s}^{-1} \AA^{-1}$.

\begin{table*}
\centering
\begin{tabular}{cccc}
\hline \hline
  & This paper & \citeauthor{nagata:2003} (\citeyear{nagata:2003}) & \citeauthor{obrien:2003} (\citeyear{obrien:2003}) \\
\hline 
$\alpha$ & $17\hour 19\min 58\secp988 \pm 0\secp008$ & $17\hour 19\min 59\secp000 \pm 0\secp 014$ & $17\hour 19\min 58\secp 994 \pm 0\secp004$ \\
$\delta$ & $-31\deg 45\amin 01\asecp21 \pm0\asecp15$  & $-31\deg 45\amin 01\asecp 2 \pm 0\asecp2$ & $-31\deg 45\amin 01\asecp 25 \pm 0.05$ \\
\hline
\end{tabular}
\caption{Summary of astrometry results (equinox J2000) for the $\xtejdsv$
NIR counterpart.}
\label{astrometry}
\end{table*}

\begin{table*}
\centering
\begin{tabular}{ccccccccc}
\hline \hline
Date & MJD & B & V & R & I & J & H & $K_{s}$ \\
\hline 
2003 02 28 & 52698 & & & & & $17.47\pm0.05$ & - & $16.00\pm0.06$ \\
2003 04 24 & 52753 & $>23.2\pm0.4$ & $>23.1\pm0.4$ & $21.5\pm0.3$ & $20.6\pm0.1$ & $17.66\pm0.05$ & $16.99\pm0.07$ & $16.34\pm0.05$ \\
\hline
\end{tabular}
\caption{Apparent B, V, R, I, J, H, and $K_{s}$ 
magnitudes of $\xtejdsv$ (MJD = JD - 2400000.5).
\label{mag_optir}}
\end{table*}

\section{Constraints on the companion star spectral type} \label{companion}

We will now try to constrain the nature of the companion star, first
with a colour-magnitude diagram, then by modelling the optical and NIR
spectral energy distribution (SED).

	\subsection{Colour-magnitude diagram}

$\xtejdsv$ optical and NIR magnitudes allow us to constrain the
nature of the binary system by comparing its absolute magnitudes with
those of well-determined spectral type stars 
(see, e.g., \citeauthor{chaty:2002b} \citeyear{chaty:2002b}). For
this purpose, we use template absolute magnitudes related to spectral
types (taken from \citeauthor{ruelas-mayorga:1991} \citeyear{ruelas-mayorga:1991} for the NIR and \citeauthor{cox:2000} \citeyear{cox:2000}
for the optical, respectively).
The conversion of apparent magnitudes $m$ to
absolute magnitudes $M$ depends on both distance $d$ and interstellar
absorption $A_{v}$, via: $M=m+5-5log d(pc)-A_{v}$.
Concerning the interstellar absorption, we have three different
estimates: First, {\it XMM-Newton} spectroscopy obtained in February 2003
gave $\nh = 1.24 \pm 0.02 \times 10^{22} \cmmoinsdeux$
(\citeauthor{cadolle-bel:2004} \citeyear{cadolle-bel:2004}). This column density corresponds to an absorption
of $A_{v}=6.9 \mags$ using the relation $\Av=5.59 \times 10^{-22} \nh$
(\citeauthor{predehl:1995} \citeyear{predehl:1995}). Second, \citeauthor{nagata:2003} (\citeyear{nagata:2003}) obtained $A_{v}= 8$ 
by assuming a high temperature blackbody emission just after the X-ray
burst. They also noted that extinction derived from 
the 2MASS survey is $A_{v} \sim 6$.

Here, we will use only one observing epoch, on  April 24, 2003,
when the source is fainter, to minimise the accretion disc
contribution in the observed NIR flux. Even if the object was still
far from quiescence, in this way, we determine a lower limit 
for the companion star spectral type 
by assuming that the accretion disc emission reddens the NIR flux.

\begin{figure}
\centering
\includegraphics[width=6cm,angle=90]{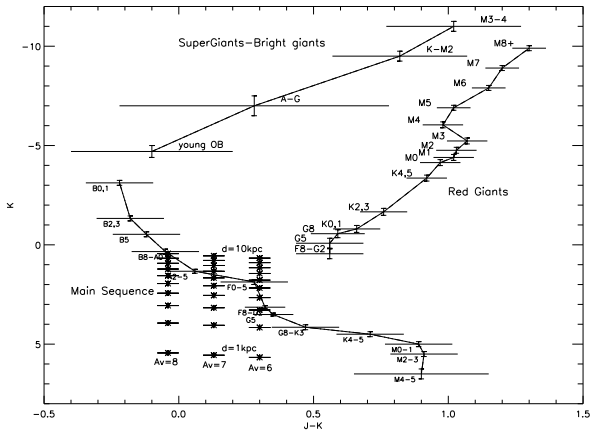} 
\caption{Colour-magnitude diagram showing characteristic absolute
magnitudes of various spectral types and $\xtejdsv$ absolute
magnitudes computed for interstellar absorption $\Av$ between 6 and
$8 \mags$ ('$\ast$' from right to left) and distance between 1 and
10 kpc ('$\ast$' from bottom to top). We used the  April 24, 2003,
observations, when the object was fainter, to reduce the
accretion disc contribution in the NIR flux. 
From this CMD we can conclude that $\xtejdsv$ is an 
intermediate mass X-ray binary located at a distance
between 3 and 10 kpc, and that the companion star is a main sequence star
of spectral type between late B and early G (see text for more details).
}
\label{CMD}
\end{figure}

To derive the possible spectral types, we computed the absolute
magnitudes of $\xtejdsv$, taking various distances $d$ and absorption in
the visible $A_{v}$. The results are reported with '$\ast$' 
in the ($J-Ks$, $Ks$) colour-magnitude diagram (CMD) presented in Fig. 
\ref{CMD}. The distance was computed between 1 and 10 kpc 
(from bottom to top, respectively) and the absorption between 6 and 8 magnitudes
(from right to left, respectively).

From this CMD, we first derive that, in any case, the companion star must
 belong to the main sequence.  Furthermore, if the interstellar
absorption is high, $\Av \sim 8 \mags$, the spectral type would be
between late B and early A, and the source far away, between 6 and 10
kpc.  With an intermediate value of the interstellar absorption,
$\Av \sim 7 \mags$, the spectral type would be between late A and
early F, and the distance between 5 and 7 kpc.  Finally, with a small
interstellar absorption, $\Av \sim 6 \mags$, the spectral type would
be between late F to early G, and the distance between 3 and 6 kpc.
Therefore, we can conclude that, for an absorption between
6 and 8 magnitudes, the $\xtejdsv$ companion star is a main sequence star
of spectral type between late B and early G, located at a distance
between 3 and 10 kpc. This estimate of distance makes the source
 closer than suggested by \citeauthor{nagata:2003} (\citeyear{nagata:2003}): 
it is therefore possible that the source is not located in the Galactic bulge.
We point out that from this analysis, 
$\xtejdsv$ can be added to the list of intermediate mass X-ray binaries, 
like, e.g., $\vqsqusgr$ (\citeauthor{chaty:2003a} \citeyear{chaty:2003a}).

	\subsection{Temperature and interstellar absorption}

We now try to further constrain the absorption and spectral types by
modelling $\xtejdsv$'s SED by an
absorbed blackbody corrected from interstellar absorption. For each
temperature, we calculate the associated blackbody spectrum that we
multiply by the transmittance filter. Taking the interstellar
absorption in the visible, we then derive absorption values for
each wavelength, using \citeauthor{cardelli:1989} (\citeyear{cardelli:1989}), 
and compute the magnitudes in each band.

\begin{figure}
\centering
\includegraphics[width=6cm,angle=90]{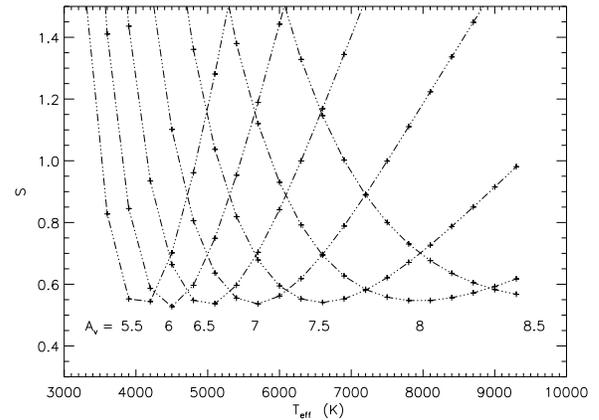} 
\caption{Minimisation of the discrepancy between apparent and computed 
magnitudes, versus the blackbody effective temperature and
interstellar absorption in the visible, given in magnitudes below each
curve.}
\label{calcspec}
\end{figure}

We now want to compare these RIJHK magnitudes, computed for
different effective temperatures and interstellar absorption, with
apparent magnitudes.
%
To do this, we calculate the discrepancy in each band 
by taking the $K_{s}$ apparent magnitude and computing
 the quantity
\[S = (R-Robs)^2+(I-Iobs)^2+(J-Jobs)^2+(H-Hobs)^2\] 
for the R-, I-, J-, and H-bands,
for all considered temperatures and absorptions.  In
Fig. \ref{calcspec}, we show the various curves of $S$ versus
effective temperatures $T$ (varying between 3300 and 9300 K)
parameterized by absorption $\Av$ (varying between 5.5 and 8.5
magnitudes).

We find that the best convergence, minimizing S, is found towards the 
values of the parameters ($A_{V} \sim 6$, $T \sim 4300 K$), 
corresponding to an $\xtejdsv$ companion star
spectral type of late F -- early G; ($A_{V} \sim 7$, $T \sim 5600 K$),
corresponding to late A -- early F; and finally ($A_{V} \sim 8$, $T
\sim 8100 K$), corresponding to late B and early A. This is, therefore,
consistent with the results given by the CMD.

\section{Evolution of $\xtejdsv$ SED} \label{SED}

\begin{figure}
\centering
\includegraphics[width=6cm,angle=90]{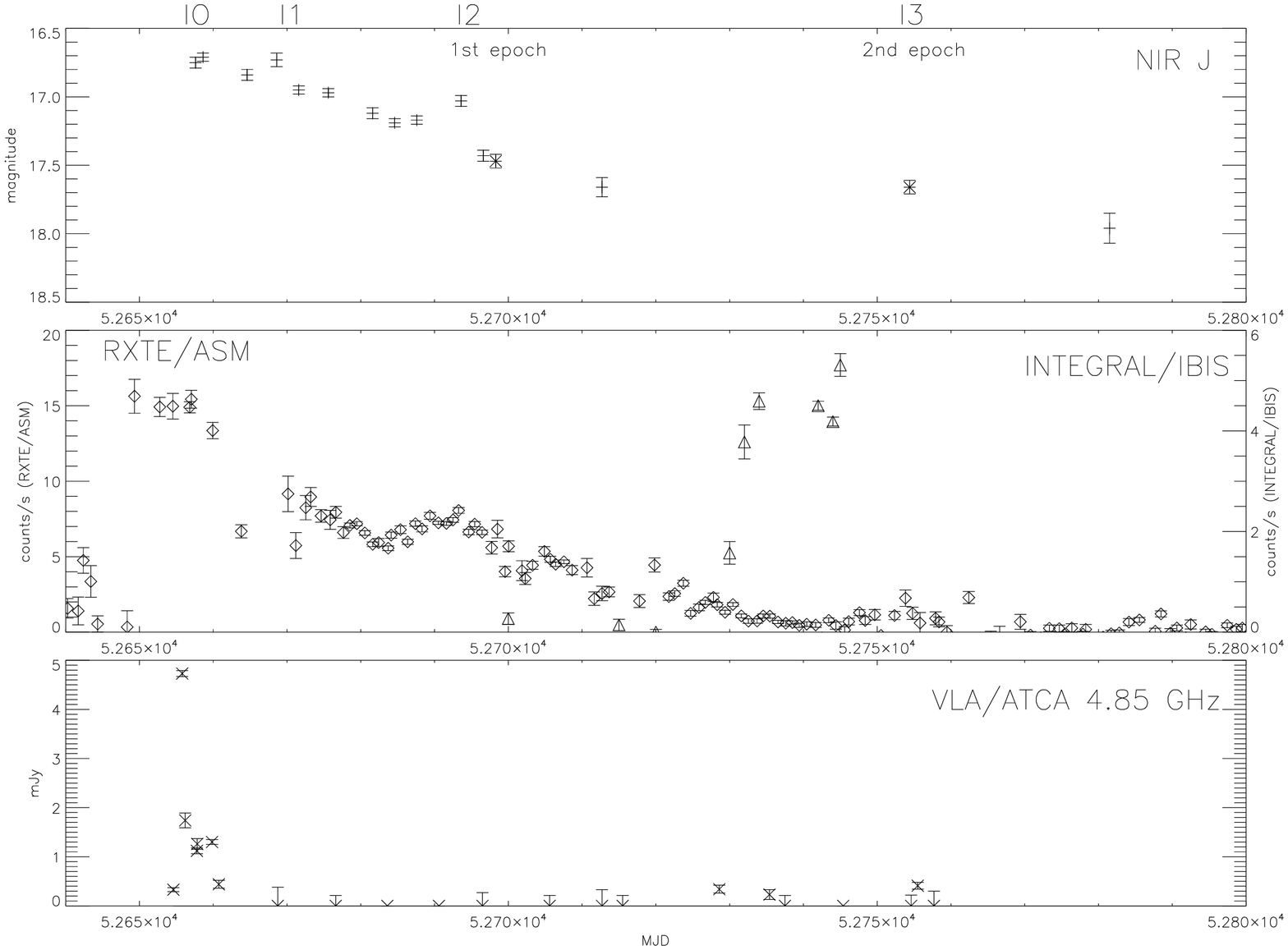}
\caption{Multi-wavelength light curve showing the outburst of $\xtejdsv$
and its transition from high-soft to low-hard state.
Top panel: NIR observations ('+': \citeauthor{nagata:2003} \citeyear{nagata:2003}, '$\ast$': this paper);
Middle panel: high-energy observations 
('$\diamond$': {\it RXTE/ASM}, '$\triangle$': {\it INTEGRAL/IBIS} \citeauthor{cadolle-bel:2004} \citeyear{cadolle-bel:2004});
Bottom panel: radio VLA/ATCA observations ('x': \citeauthor{brocksopp:2005} \citeyear{brocksopp:2005}).
We indicated at the top the times when the main outburst (O)
and the three secondary outbursts (1, 2, 3 respectively) occurred.
MJD = JD - 2400000.5}
\label{lc}
\end{figure}

In Fig. \ref{lc}, we show the overall light curve of $\xtejdsv$ from 
mid-January to the end of August 2003.
First we can see that the light curve  grossly has the form
of a Fast-Rise Exponential-Decay
light curve (so-called FRED), but superimposed on this FRED, the source 
exhibits a complex behaviour:
after the main outburst on January 16 (MJD 52656), 
we can see both in X-rays and in NIR 
a secondary outburst on  January 29 (MJD 52669), 
then a second one on February 22 (MJD 52693),
and finally a third one on April 24, 2003 (MJD 52753), 
exactly at the time of our second epoch NIR observations.
The main outburst and the last event are also associated
with radio outbursts, and therefore with ejection events.
We indicate the time of these events in Fig. \ref{lc}
by O, 1, 2, and 3, respectively.
There are other sources that exhibit clear secondary maxima in their
X-ray light curves, such as A0620-00, GS 1124-68, GRO J0422+32 
(\citeauthor{chen:1997} \citeyear{chen:1997}), 4U 1543-47 (\citeauthor{buxton:2004} \citeyear{buxton:2004}), 
and XTE J1550-564 (\citeauthor{jain:2001} \citeyear{jain:2001}).
However, $\xtejdsv$ seems to be the second source after A0620-00
to exhibit clear multi-secondary maxima in the optical/NIR,
correlated with the X-rays, as seen in Fig. \ref{lc}.

We also report the SED of $\xtejdsv$ in Fig. \ref{sed} for the two
observing epochs 
('+' for  February 28, 2003, and '$\ast$' for  April 24, 2003, respectively), 
where we put together ESO/NTT optical/NIR observations from this paper, 
{\it INTEGRAL/IBIS} high-energy observations (\citeauthor{cadolle-bel:2004} \citeyear{cadolle-bel:2004}), 
and ATCA/VLA radio data (\citeauthor{brocksopp:2005} \citeyear{brocksopp:2005}). 
We will now describe the light curve and both SEDs and analyse $\xtejdsv$'s
evolution between these two observing epochs.

	\subsection{The  February 2003 epoch: high-soft state}

Observations obtained with {\it XMM-Newton}, {\it INTEGRAL/IBIS}, and 
{\it Rossi-XTE} in  February 2003 showed that the source spectrum was very
soft (power law index of $\Gamma = 2.7$): 
while the 2-12 keV flux was $\sim 100$ mCrab, {\it IBIS}
detected a source at only $\sim 2.1$ mCrab in the 20-120 keV band 
 (\citeauthor{cadolle-bel:2004} \citeyear{cadolle-bel:2004}).
$\xtejdsv$ was not detected in radio 
($\leq 0.27$ mJy at 4.8 GHz, \citeauthor{brocksopp:2005} \citeyear{brocksopp:2005}),
which is consistent with typical high-soft state (see, e.g., \citeauthor{chaty:2003b} \citeyear{chaty:2003b}).
This is corroborated by the analysis of $\xtejdsv$ SED 
from  February 28, to March 2, 2003 (reported 
with '+' in Fig. \ref{sed}), where we notice the characteristic shape 
of the accretion disc emission in X-rays
(strong flux and soft spectrum), and the absence of radio emission.

	\subsection{The  April 2003 epoch: low-hard state}

\begin{figure}
\centering
\includegraphics[width=6cm,angle=90]{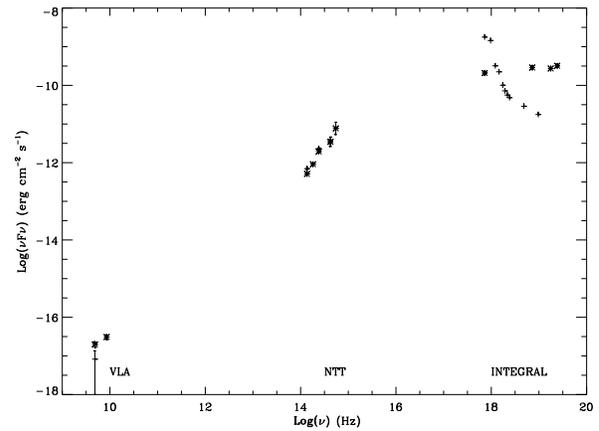}
\caption{Spectral Energy Distribution of $\xtejdsv$ during
the two observing epochs: the  February 28, 2003, data are represented by
'+', and the  April 24, 2003, by '$\ast$'.  The observations were
taken nearly simultaneously with the VLA in the radio  (\citeauthor{brocksopp:2005} \citeyear{brocksopp:2005}), 
the NTT/EMMI and SOFI in the optical and NIR (this paper), 
and {\it INTEGRAL/IBIS} in the high-energy (\citeauthor{cadolle-bel:2004} \citeyear{cadolle-bel:2004}). For
the first epoch, the VLA data were taken on  February 26, 2003, NTT/SOFI
data on  February 28, 2003, and {\it INTEGRAL/IBIS} data on  February 28
-- March 2, 2003. For the second epoch, VLA data were taken on  April
26, 2003, NTT/SOFI data on  April 24, 2003, NTT/EMMI on April 27, 2003, and
{\it INTEGRAL/IBIS} data on  February 28 -- March 2, 2003. The 
February 28, 2003, observations correspond to the high-soft state: high and
soft X-ray flux and no radio emission. On the contrary, the 
April 24, 2003, observations correspond to the low-hard state: low and hard
X-ray flux and detection of radio emission. Optical and
NIR fluxes were de-reddened, assuming an interstellar absorption in the
visible of $\Av = 7 \mags$. It is remarkable that they remain the same 
in both states.}
\label{sed}
\end{figure}

On the other hand, after  March 25, 2003 (MJD 52723),
 $\xtejdsv$ hard X-ray flux as seen by {\it INTEGRAL/IBIS}
increased by a factor of 100 with respect 
to the high-soft state (\citeauthor{cadolle-bel:2004} \citeyear{cadolle-bel:2004}), while the soft X-ray flux
({\it RXTE/ASM}) remained constant, as seen on the 
light curve of Fig. \ref{lc}.
Besides, we also observed a radio burst (\citeauthor{brocksopp:2005} \citeyear{brocksopp:2005}), 
simultaneous with an X-ray burst and an increase in the NIR flux.
Therefore, $\xtejdsv$ seems to have entered a transition towards a low-hard 
state in-between these 2 observing epochs, as suggested by \citeauthor{goldoni:2003} (\citeyear{goldoni:2003}).
This is confirmed by the analysis of the SED (data of this second epoch 
are reported with '$\ast$' in Fig. \ref{sed}): 
the source shows all the usual signs of the low-hard state.
Firstly, we immediately notice that the source hardened
in the high-energy domain with a high-energy power law index 
of $\Gamma = 1.8$.
Secondly, the radio emission is usually interpreted in this state as
synchrotron emission emanating from a jet. 
From the SED, we can derive the power law index $\alpha$ 
(in $S_\nu \propto \nu^\alpha$):
$\sim -0.3$ in the radio and $\sim 1.6$ in the optical/NIR.
Therefore, the extrapolation of the radio flux towards the optical/NIR domain 
is significantly fainter than the observed optical/NIR flux.
This strongly suggests that the synchrotron emission from the jet is
contributing only for a small part, if any, in the NIR emission.
It seems also likely that there is no contribution from the
accretion disk in the optical/NIR domain, since the slope in the NIR
and optical remarkably remains 
the same in both observing epochs, while the X-ray decreased at
the same time. Therefore, the NIR emission is dominated 
by the contribution of the companion star, 
which is consistent with $\xtejdsv$ being an intermediate mass X-ray binary.

To further understand this source, it would
be useful i) to get spectroscopic observations of $\xtejdsv$ in
quiescence to better characterise the
companion star when the photospheric flux of the star dominates
and ii) to observe the radial velocity of the binary system to
derive the mass function and orbital parameters. With these 
parameters we will be able 
to further analyse its SED (as in, e.g., \citeauthor{chaty:2003b} \citeyear{chaty:2003b} 
with the source $\xtejodh$).

%

\section{Conclusions}

We have reported optical and NIR observations of the X-ray binary
$\xtejdsv$, taken as Target of Opportunity observations following the
 January 2003 outburst of this source. 
By performing accurate astrometry, we  discovered the optical counterpart 
in the R-band (R $\sim 21.5$) and confirmed the near-infrared counterpart. 
From photometric observations,
analysis of a colour-magnitude diagram, and a basic modelling of its SED,
we found that, for an absorption between 6 and 8 magnitudes,
$\xtejdsv$ is likely to be an intermediate mass X-ray binary, hosting 
a black hole and a main sequence star of spectral type between late B
and early G, located at a distance between 3 and 10 kpc.
We also analysed the $\xtejdsv$ X-ray and near-infrared 
light curves: this source exhibited three secondary outbursts,
and our second set of observations took place simultaneously with the third
 one. Comparing the SEDs during and after its outburst, we confirm 
the change of state of this source, from high-soft to low-hard state.

\begin{acknowledgements}
SC thanks the ESO staff and especially Malvina Bill\`eres, C\'edric Foellmi, 
Lisa Germany, Olivier Hainaut, Gaspare Lo-Curto, and
Emanuela Pompei for performing the ToO observations.
We are grateful to Marion Cadolle-Bel and St\'ephane Corbel 
for making IBIS and ATCA data, respectively, available to us 
prior to publication.
IRAF is distributed by the National Optical Astronomy Observatories,
    which are operated by the Association of Universities for Research
    in Astronomy, Inc., under a cooperative agreement with the National
    Science Foundation.
This research has made use of NASA's Astrophysics Data System 
Bibliographic Services.
{\it XTE} Results were provided by the ASM/RXTE teams at MIT and at the RXTE SOF 
and GOF at NASA's GSFC.
\end{acknowledgements}

\bibliographystyle{aa} 
\bibliography{science}

\end{document}